\documentclass[aps, prl, amsmath, amssymb, english, twocolumn,reprint,floatfix, superscriptaddress]{revtex4-2}

\usepackage[english]{babel}
\usepackage{graphicx}
\usepackage{verbatim}

\usepackage{epsfig}
\usepackage{epstopdf}
\usepackage{amsfonts}
\usepackage{amsthm}
\usepackage{amsmath}
\usepackage{amssymb}
\usepackage{color}
\usepackage[usenames,dvipsnames,svgnames,table]{xcolor}
\usepackage{hyperref} 
\hypersetup{pdfpagemode=FullScreen,colorlinks=true,linkcolor=NavyBlue,citecolor=BrickRed}
\usepackage{makeidx}
\usepackage{pifont}

\usepackage{xr}
\usepackage{color}
\usepackage{grffile}

\usepackage[ruled]{algorithm2e}
\usepackage[noend]{algpseudocode}
\usepackage{multirow}
\usepackage{type1cm} 
\usepackage{dcolumn}
\usepackage{bm}

\begin{document}

\title{Prediction of several Co-based La$_3$Ni$_2$O$_7$-like superconducting materials}

\author{Jing-Xuan Wang}
\thanks{These authors contributed equally to this work.}
\affiliation{School of Physics and Beijing Key Laboratory of Opto-electronic Functional Materials $\&$ Micro-nano Devices, Renmin University of China, Beijing 100872, China}\affiliation{Key Laboratory of Quantum State Construction and Manipulation (Ministry of Education), Renmin University of China, Beijing 100872, China}
\author{Yi-Heng Tian}
\thanks{These authors contributed equally to this work.}
\affiliation{School of Physics and Beijing Key Laboratory of Opto-electronic Functional Materials $\&$ Micro-nano Devices, Renmin University of China, Beijing 100872, China}\affiliation{Key Laboratory of Quantum State Construction and Manipulation (Ministry of Education), Renmin University of China, Beijing 100872, China}
\author{Jian-Hong She}
\thanks{These authors contributed equally to this work.}
\affiliation{School of Physics and Beijing Key Laboratory of Opto-electronic Functional Materials $\&$ Micro-nano Devices, Renmin University of China, Beijing 100872, China}\affiliation{Key Laboratory of Quantum State Construction and Manipulation (Ministry of Education), Renmin University of China, Beijing 100872, China}
\author{Rong-Qiang He}\email{rqhe@ruc.edu.cn}\affiliation{School of Physics and Beijing Key Laboratory of Opto-electronic Functional Materials $\&$ Micro-nano Devices, Renmin University of China, Beijing 100872, China}\affiliation{Key Laboratory of Quantum State Construction and Manipulation (Ministry of Education), Renmin University of China, Beijing 100872, China}
\author{Zhong-Yi Lu}\email{zlu@ruc.edu.cn}\affiliation{School of Physics and Beijing Key Laboratory of Opto-electronic Functional Materials $\&$ Micro-nano Devices, Renmin University of China, Beijing 100872, China}\affiliation{Key Laboratory of Quantum State Construction and Manipulation (Ministry of Education), Renmin University of China, Beijing 100872, China}\affiliation{Hefei National Laboratory, Hefei 230088, China}

\date{\today}

\begin{abstract}

High-temperature superconductivity has been found in Fe-, Ni-, and Cu-based compounds but has remained elusive in Co-based materials. The recent discovery of superconductivity in pressurized bilayer nickelate La$_3$Ni$_2$O$_7$ has renewed interest in related layered systems. Here, we predict several Co-based analogs that may realize similar physics. Electron doping of the high-pressure bilayer cobaltate La$_3$Co$_2$O$_7$ yields LaTh$_2$Co$_2$O$_7$, La$_3$Ni$_2$O$_5$Cl$_2$, and La$_3$Ni$_2$O$_5$Br$_2$, which exhibit closely related crystal structures and strongly correlated electronic states. Random-phase-approximation calculations reveal $s$-wave as the leading pairing symmetry in these compounds.

\end{abstract}


\maketitle

{\it Introduction}. Since the discovery of cuprate superconductors, their unconventional properties have attracted sustained interest~\cite{cw1987science,schilling1993nature,putilin1993nature,park1995jacs}. Subsequent discoveries of high-temperature superconducting phases in Fe- and Ni-based compounds have further highlighted the role of strong correlations in 3$d$ transition-metal systems~\cite{iron-based2008kamihara,rotter2008prl,johnston2010puzzle,li2020superconducting,li2019superconductivity,sun2023nature,zhu2024superconductivity}. Nickelate compounds were theoretically predicted to exhibit superconductivity by analogy with cuprates, and several materials have since been confirmed experimentally~\cite{botana2020PRX,poltavets2009prl,poltavets2010prl,sakakibara2020prl}. The successful extension of superconductivity from cuprates to nickelates provides a promising avenue for the exploration of other transition-metal-based superconductors. Recently, pressurized bilayer nickelate La$_3$Ni$_2$O$_7$ (LNO) has been established as a high-temperature superconductor with $T_c \approx 80$ K, driven by strong electronic correlations and associated instabilities~\cite{sun2023nature}. Motivated by this, it is natural to investigate isoelectronic systems beyond nickel—such as cobaltates, where Co exhibits tunable spin states—as potential platforms for high-$T_c$ superconductivity.

The physics of LNO is characterized by two key features: a bilayer crystal structure and a distinctive electronic structure, where strong interlayer hopping drives the nearly half-filled Ni $d_{z^2}$ orbitals to form bonding and antibonding bands~\cite{luo2023prl,zhang2023prbl,Viktor2023prl}. A prevailing view is that the strong interlayer antiferromagnetic correlation of the $d_{z^2}$ orbitals play a central role in superconducting pairing~\cite{Yang2023PRB,ShenCPB2023,SakakibaraPRL2024,luo2024npj,zheng2025prb,shen2025prb,chen2024prb}. In addition, Hund's-coupling correlations are widely recognized as a key ingredient in these nickelate superconductors~\cite{lu2024prl,qu2024prl,oh2023prb,tian2024prb,chen2024prb,ouyang2024prb,cao2024prb}. A computational study has highlighted the magnitude of the Ni local moment as a critical parameter~\cite{1412-nfzm}, finding that superconductivity appears only within a narrow window of the Ni magnetic moment, between robust magnetic order and a Fermi-liquid state. These observations suggest that designing materials isostructural to LNO, while retaining the above electronic characteristics, is a promising route to expanding the family of high-temperature superconductors. Given that Co is adjacent to Ni in the periodic table, cobaltates are natural candidates for trying this strategy.

The discovery of cobaltate superconductivity dates back to 2003, when superconductivity with a transition temperature of 3 K was first reported in the layered cobaltate Na$_x$CoO$_2$$\cdot$H$_2$O~\cite{takada2003superconductivity}. In this compound, the pairing mechanism remains unclear~\cite{mazin2005critical}, and single crystals are difficult to synthesize~\cite{lynn2003prb}. Subsequently, the field expanded to include materials structurally analogous to iron pnictides, such as the ``122"-type LaCo$_2$B$_2$ ($T_c \approx 4$~K)~\cite{mizoguchi2011prl} and the ``111"-type LaCoSi ($T_c \approx 4$~K)~\cite{he2021inorgchem}. More recently, Na$_2$CoSe$_2$O has been synthesized and found to exhibit a superconducting transition temperature of 6.3~K~\cite{chengJACS2024}. Although this material shows an exceptionally high upper critical field, recent work suggests that it likely hosts conventional superconductivity~\cite{wu2025weak}. Ref.~\cite{HuAPS2021} predicts Co-based superconducting platforms with Co$^{2+}$ in tetrahedral and trigonal-bipyramidal environments, yet the predicted superconducting phases have not been realized experimentally. Despite these advances, high-temperature unconventional superconductivity has not yet been established in cobaltates, leaving a conspicuous gap relative to other 3$d$ transition-metal oxides. In this broader search for new Co-based superconducting or correlated-electron platforms, La$_3$Co$_2$O$_7$ (LCO) is particularly intriguing: it has been synthesized previously as an $n=2$ member of the La–Co–O Ruddlesden--Popper series for catalytic and oxygen-electrode applications~\cite{huan2019intrinsic,nagy2025based}, but its low-temperature electronic and magnetic properties remain essentially unexplored.


In this Letter, we theoretically identify bilayer LNO-like cobaltates as promising candidates for high-temperature superconductivity by electron-doping the high-pressure phase of LCO. By analogy with LNO, we use density functional theory plus dynamical mean-field theory (DFT+DMFT) and random-phase approximation (RPA) methods~\cite{graserNJP2009,kemperNJP2010} to investigate the electronic structure, electronic correlations, and superconductivity of LaTh$_2$Co$_2$O$_7$ (LCO-Th) and La$_3$Co$_2$O$_5$Cl$_2$ (LCO-Cl). Their band structures closely resemble those of LNO, and their Co $e_g$ orbitals exhibit correlations and local moments comparable to those in La$_3$Ni$_2$O$_7$. The Th- and Cl-doped LCO compounds display strongly correlated electronic structures, characterized by strong Hund's coupling and large mass enhancements, which favor the emergence of superconductivity. These results suggest that pressure and doping may induce high-temperature superconductivity in LCO, in close analogy with the nickelates.


\begin{figure}[htb]
  \centering
  \includegraphics[width=8.6cm]{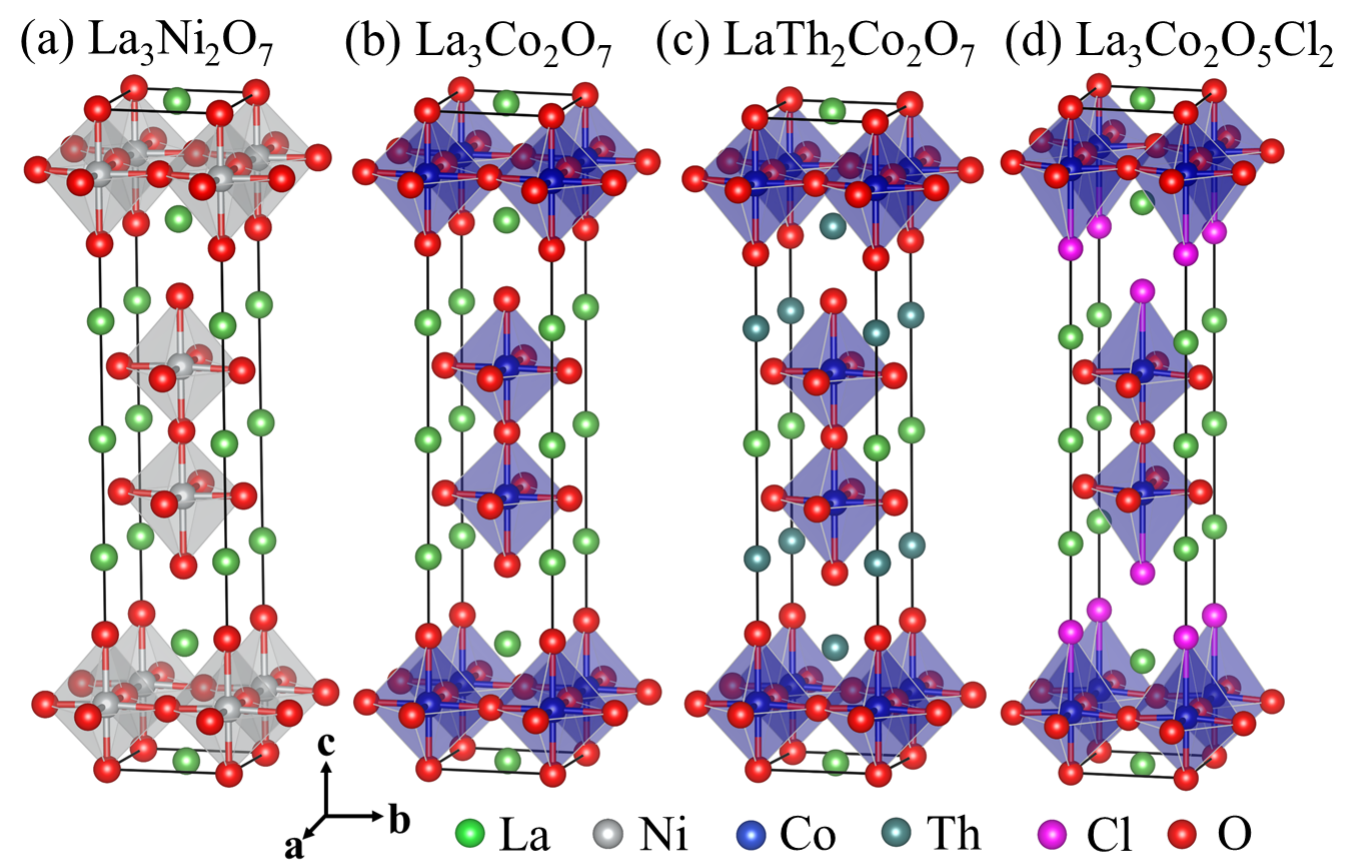}
  \caption{Crystal structure of (a) LNO, (b) LCO, (c) LCO-Th, and (d) LCO-Cl at high pressure.}
  \label{fig:structure}
\end{figure}

{\it Similar crystal structures}. 
To facilitate a direct comparison with the superconducting phase of LNO at a pressure of 29.5~GPa, we optimized the structure of LCO at 25~GPa and obtained comparable lattice constants, as summarized in Table~S1 of the Supplemental Material (SM)~\cite{SM}. The space group of LCO is $I4/mmm$, with lattice parameters $a = 3.703$ \text{\AA} and $c = 19.174$ \text{\AA} at 25 GPa. In addition, both the bond lengths and bond angles of the transition-metal--O bonds in the two materials are comparable. To confirm the dynamical stability, we calculated the phonon spectra within DFT and found no imaginary phonon frequencies, which indicates that the structure is dynamically stable (see Fig.~S1 in the SM). 

The Co ion in LCO has a formal valence of $+2.5$ with a 3$d^{6.5}$ electronic configuration, whereas the Ni ion in LNO has a 3$d^{7.5}$ configuration. To realize a similar 3$d^{7.5}$ configuration for Co, we introduce electron doping by substituting tetravalent Th cation at the La site or monovalent Cl anion at the O site, thereby reducing the Co valence from $+2.5$ to $+1.5$. In LCO-Th, the in-plane Co--O bond lengths within the CoO$_6$ octahedra closely match those in LNO, while the apical Co--O bonds are elongated by approximately $2\%$ relative to those in LNO. In LCO-Cl, the in-plane Co--O bond lengths remain similar to those in LNO, whereas the Co--Cl bond is significantly longer (by about $25\%$) than the corresponding Ni--O bond in LNO, which we attribute to the smaller ionic charge of the Cl$^-$ anion compared with O$^{2-}$. Further structural details are provided in the SM.

\begin{figure}[tb]
  \centering
  \includegraphics[width=8.6cm]{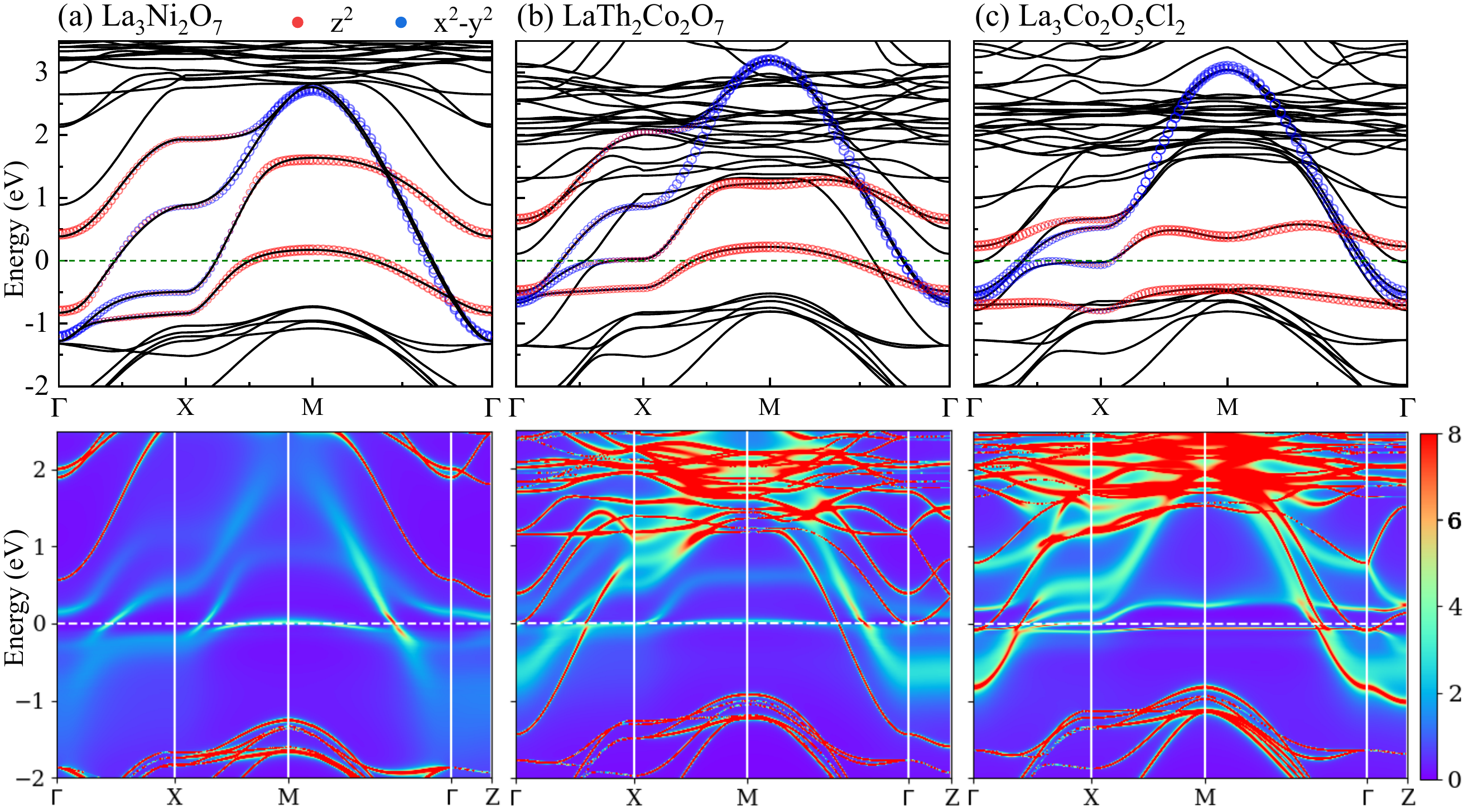}
  \caption{Band structures of LNO, LCO-Th, and LCO-Cl. The DFT bands are shown in the upper panels. The band structure of the fitted bilayer two-orbital tight-binding models is superposed, where the orbital weights are represented by the size of the colored circles. The momentum-resolved spectral functions $A(\mathbf{k},\omega)$ obtained by DFT+DMFT at 290~K are displayed in the lower panels. The dashed green and blue lines at 0 eV denote the Fermi level.}
  \label{fig:band}
\end{figure}

{\it Similar electronic structures}. Identifying favorable electronic structures is central to predicting new superconductors. Fig.~\ref{fig:band} presents the band structures of LNO, LCO-Th, and LCO-Cl from DFT, together with the momentum-resolved spectral functions from DFT+DMFT. Among the three materials, the Ni/Co $e_g$ bands are similar: they have a bandwidth of approximately 3.8~eV, spanning from $-1.2$ to 2.6~eV for Ni and from $-0.8$ to 3.0~eV for Co. Unlike in LNO, the bands near the Fermi level in Th- and Cl-doped cobaltates are less orbital-pure, with noticeable contributions from other orbitals. In LNO, the La 5$d$ bands extend down to about 0.9 eV above the Fermi level, whereas in LCO-Th the Th 4$f$ bands cross the Fermi level and extend down to about $-$1.4 eV, forming a pocket at the $\Gamma$ point dominated by Th orbitals. Similarly, in LCO-Cl the La 5$d$ bands appear near the Fermi level. The partial occupancy of these La 5$d$ or Th 4$f$ bands is analogous to the self-doping effect in LaNiO$_2$~\cite{anisimov1999prb,botana2020PRX,hepting2020electronic} and in rare-earth-element-doped nickelate~\cite{electronic2019prb,hausoel2025superconducting,liu2025superconductivity,lv2025growth,zhou2025ambient} and cuprate superconductors~\cite{tokura1989superconducting,RevModPhys.82.2421,rybicki2016perspective}, and thus is not expected to suppress superconductivity.

In the lower panels of~Fig.~\ref{fig:band}, the momentum-resolved spectral functions calculated by DFT+DMFT at 290~K show that the $e_g$ bands become substantially broadened, indicating strong correlation. The correlation-induced renormalization further narrows the $e_g$ bandwidths in both LCO-Th and LCO-Cl, resulting in nearly flat bands at the Fermi level. These features closely resemble those found in LNO.


\begin{figure}[tb]
  \centering
  \includegraphics[width=7.6cm]{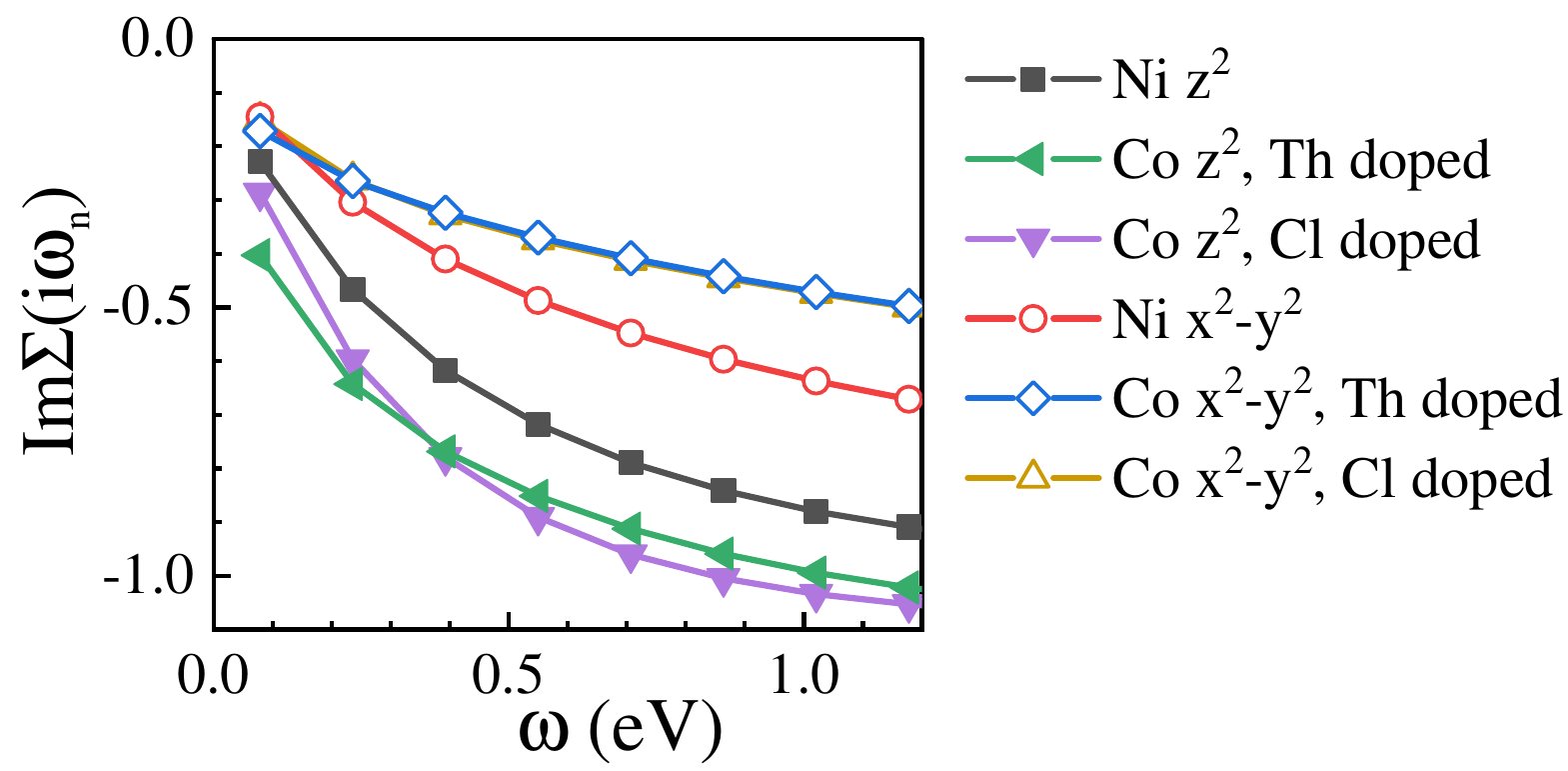}
  \caption{The imaginary parts of the self-energies $\rm{Im}\Sigma(i\omega_n)$ at the Matsubara axis for LNO, LCO-Th, and LCO-Cl at 290~K.}
  \label{fig:self-energy}
\end{figure}


\begin{table}[t]
  \centering
  \renewcommand\arraystretch{1.5}
  \caption{Local orbital occupation number $N_d$ and effective mass enhancement $m^*/m$ of the Ni and Co atoms in LNO and LCO with Th, Cl and Br doped at high pressure for the $d_{z^2}$ and $d_{x^2-y^2}$ orbitals at 290~K.}
  \label{tab2}
  \begin{tabular*}{8.6cm}{@{\extracolsep{\fill}} lcccc}
    \hline\hline
    \multirow{2}{*}{\centering } & 
    \multicolumn{2}{c}{$N_d$} & 
    \multicolumn{2}{c}{$m^*/m$} \\
    & $d_{z^2}$ & $d_{x^2-y^2}$ & $d_{z^2}$ & $d_{x^2-y^2}$ \\
    \hline
    Ni in La$_3$Ni$_2$O$_7$ & 1.136 & 1.056 & 2.52 & 2.02 \\
    Co in LaTh$_2$Co$_2$O$_7$& 0.938 & 0.748 & 5.81 & 4.02\\
    Co in La$_3$Co$_2$O$_5$Cl$_2$ & 1.004 & 0.726 & 3.33 & 2.93\\
    Co in La$_3$Co$_2$O$_5$Br$_2$ & 1.001 & 0.719 & 2.13 & 2.89\\
    \hline\hline
  \end{tabular*}
\end{table}

\begin{table}[t]
  \centering
  \renewcommand\arraystretch{1.5}
  \caption{The weights (\%) of the Ni- and Co-$e_g$ orbital local multiplets for LNO and LCO with Th, Cl and Br doped respectively calculated by DFT+DMFT at 290 K. The good quantum numbers $N_\Gamma$ and $S_z$ denote the total occupancy and total spin of the Ni- and Co-$e_g$ orbital, which are used to label different local spin states.}
  \label{tab3}
  \begin{tabular*}{8.6cm}{@{\extracolsep{\fill}} lcccccccc}
    \hline\hline
    $N_\Gamma$& 0 & 1 & 2 & 2 &3 & 4&$ $\\
    $S_z$& 0 & 1/2 & 0 & 1 &1/2 & 0&$\sqrt{\langle S_z^2 \rangle}$ \\
    \hline
    Ni in La$_3$Ni$_2$O$_7$& 0.0& 12.4 & 23.6 & 33.3 & 28.1& 2.3&0.658\\
    Co in LaTh$_2$Co$_2$O$_7$& 2.5& 36.2& 21.4 & 30.3 & 9.2 & 0.3&0.645\\
    Co in La$_3$Co$_2$O$_5$Cl$_2$& 2.2& 33.6 & 22.7 & 30.9 & 10.2 & 0.4&0.647\\
    Co in La$_3$Co$_2$O$_5$Br$_2$& 2.4& 34.1 & 23.4 & 29.5 & 10.2 & 0.4&0.637\\
    \hline\hline
  \end{tabular*}
\end{table}


{\it Similarly strong correlations}. To investigate correlation effects, Fig.~\ref{fig:self-energy} presents the imaginary part of self-energy, Im$\Sigma(i\omega_n)$, at Matsubara frequencies for the Co $e_g$ orbitals in Th- and Cl-doped LCO, compared with the Ni $e_g$ orbitals in LNO. Im$\Sigma(i\omega_n)$ of the Co $d_{x^2-y^2}$ orbital is similar to that for the Ni $d_{x^2-y^2}$ orbital in LNO, but with smaller magnitudes. Im$\Sigma(i\omega_n)$ of the Co $d_{z^2}$ orbital exhibits a profile remarkably similar to that of the Ni $d_{z^2}$ orbital in LNO. For the Th-doped cobaltate, Im$\Sigma(i\omega_n)$ of the Co $d_{z^2}$ orbital exhibits a finite intercept, signaling non-Fermi-liquid behavior. For the Cl-doped cobaltate, Im$\Sigma(i\omega_n)$ for the Co $d_{z^2}$ orbital displays nonlinear frequency dependence at low frequencies, implying potential strange metal behavior. In addition, the magnitudes of Im$\Sigma(i\omega_n)$ for the $d_{z^2}$ orbitals are larger than those for the $d_{x^2-y^2}$ orbitals, indicating stronger correlations in the former. These features resemble those observed for the Ni in LNO.

Table~\ref{tab2} lists the occupation numbers $N_d$ and effective mass enhancements $m^*/m$ of the $e_g$ orbitals for LNO and electron-doped LCO. The Co $d_{z^2}$ orbitals are half-filled, like the Ni $d_{z^2}$ orbitals, which would be the main reason for the strong correlation. The Co $d_{x^2-y^2}$ orbitals have a lower occupancy ($\sim$ 0.73) than the Ni counterpart, however, this does not eliminate the strong correlation, implying that a $\sim$ 0.73 occupancy of the $d_{x^2-y^2}$ orbital is sufficient for the Hund's spin correlation~\cite{tian2024prb,ouyang2024prb,wjx2024prb} to exist and to play a role in the electron-doped LCO.

For the electron-doped LCO, both the low-spin state ($N_{\Gamma}=1$, $S_z=1/2$) and the high-spin state ($N_{\Gamma}=2$, $S_z=1$) exhibit substantial weights (shown in Table \ref{tab3}), signaling a doping-induced evolution of Co from a predominantly low-spin configuration to a mixed high-spin/low-spin state. Crucially, the local magnetic moments we calculate for these cobalt-based compounds---ranging from 0.637 to 0.647---fall precisely within the narrow window ($\approx$ 0.63\text{--}0.68) recently identified as optimal for high-$T_c$ superconductivity in Ruddlesden--Popper nickelates~\cite{1412-nfzm}. This close alignment provides compelling theoretical evidence that these cobaltates may similarly host strong spin fluctuations conducive to realizing high-temperature superconductivity.



\begin{table*}[t]
  \centering
  \renewcommand\arraystretch{1.5}
  \caption{Parameters of our bilayer two-orbital TB models, compared with Ref.~\cite{luo2023prl}.}
  \label{tab1}
  \begin{tabular*}{18cm}{@{\extracolsep{\fill}} lccccccccccccccccccc}
    \hline\hline
     & $t_1^x$ & $t_1^z$ & $t_2^x$ & $t_2^z$ &  $t_3^{xz}$ & $t_\perp^x$ &$t_\perp^z$ & $t_4^{xz}$ & $\epsilon^x$ &$\epsilon^z$ & $t_3^x$ & $t_3^z$ & $t_4^x$ & $t_4^z$ &  $t_5^{xz}$ & $t_5^x$ & $t_5^z$  &$t_6^x$ & $t_6^z$ \\
    \hline
    La$_3$Ni$_2$O$_7$ \cite{luo2023prl} & -.483 & -.110 & .069 & -.017 & .239 & ~.005 & -.635 & -.034 & ~.776 & ~.409  \\
    La$_3$Ni$_2$O$_7$  & -.492 & -.135 & .059 & -.011 & .253 & -.013 & -.670 & -.043 & ~.721 & ~.470 & -.001 & ~.013 & .019 & -.013 & .032 & -.050 & -.019 & -.013 & .014\\
    LaTh$_2$Co$_2$O$_7$&  -.478 & -.078 & .057 & -.001 & .196 & -.090 & -.578 & -.058 & 1.168 & ~.494&  -.001 & -.009 & .031 & ~.001 & .028 & -.028 & -.021 & -.012 &  .010\\
    La$_3$Co$_2$O$_5$Cl$_2$ & -.459 & -.026 & .112 & ~.005 & .087 & ~.017 & -.495 & ~.004 & 1.000 & -.070& ~.010 & -.005 & .001 & ~.000 & .024 & -.043 & -.021 & -.001 & .014 \\
    \hline\hline
  \end{tabular*}
\end{table*}

\begin{figure}[htb]
  \centering
  \includegraphics[width=8.6cm]{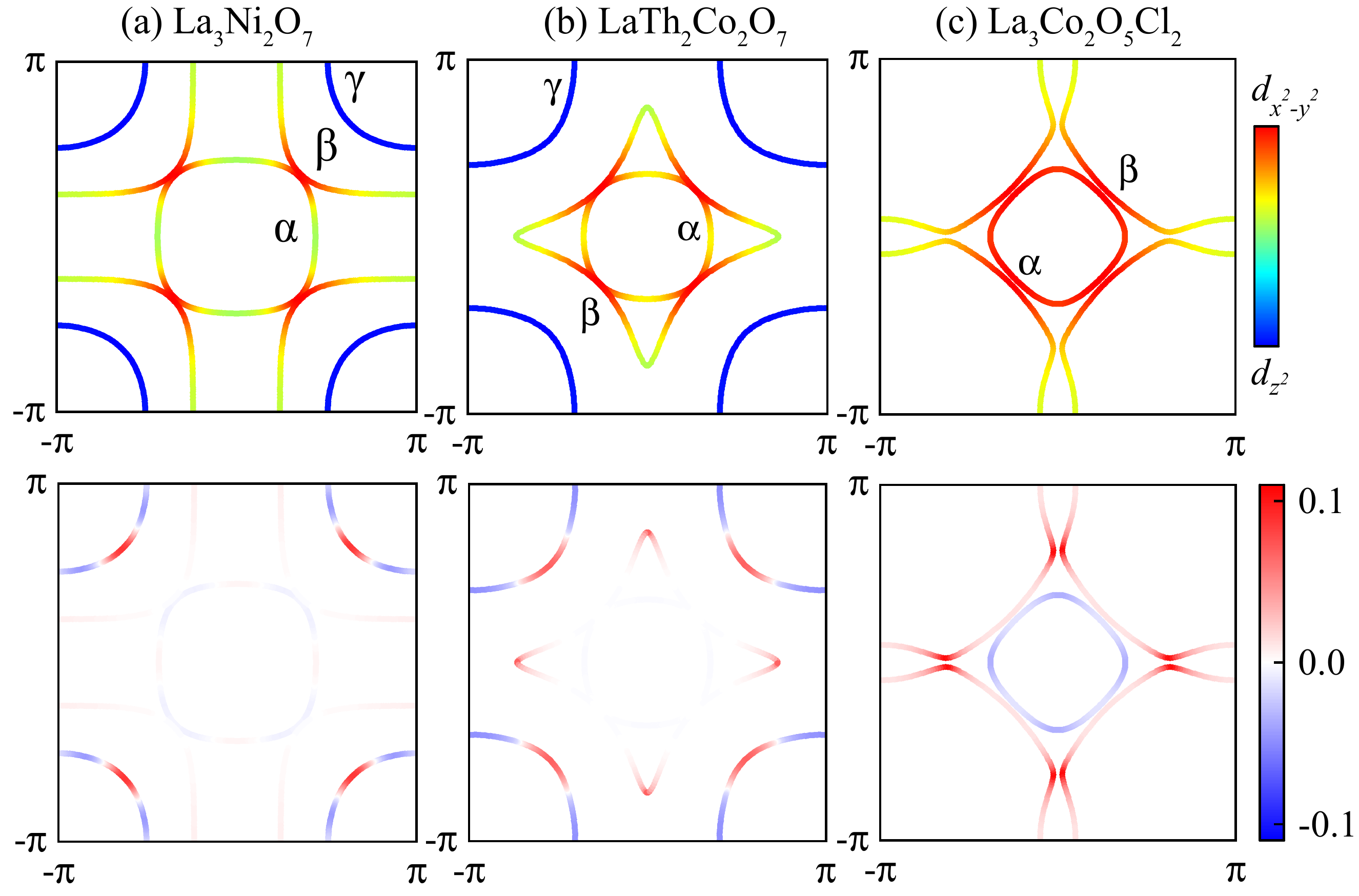}
  \caption{The upper panels show the orbital-resolved Fermi surfaces of the corresponding TB models for (a) LNO, (b) LCO-Th, and (c) LCO-Cl. The lower panels show the superconducting gap structures on the Fermi surfaces calculated with RPA on the three TB models with $U=1.4$, $0.9$, and $1$ eV, respectively.}
  \label{fig:fermisurface}
\end{figure}

{\it Tight-binding models and $s$-wave pairing symmetry}. 
To investigate superconducting properties, we construct tight-binding (TB) models by fitting to the $e_g$ bands from our DFT calculations via least-squares fitting, and then apply the RPA to determine the pairing symmetry. The onsite energies $\epsilon_{x/z}$ and hopping parameters are listed in Table~\ref{tab1} for LCO-Th and LCO-Cl, with LNO included for comparison. As shown in Fig.~\ref{fig:band}, the TB models capture the main characteristics of the Co $e_g$ bands near the Fermi level.

The upper-middle panel of Fig.~\ref{fig:fermisurface} shows the Fermi surface of the TB model for LCO-Th, which comprises three sheets ($\alpha$, $\beta$, and $\gamma$). The critical interaction strength $U_c$ at which the spin susceptibility diverges is $1.01$~eV. The leading superconducting gap function for $U=0.9$~eV belongs to the $A_{1g}$ irreducible representation, thus has $s$-wave symmetry, as shown in the lower-middle panel of Fig.~\ref{fig:fermisurface}.


The Fermi surface of the TB model for LCO-Cl, shown in the upper-right panel of Fig.~\ref{fig:fermisurface}, consists of only two sheets, $\alpha$ and $\beta$. The corresponding critical interaction strength is $U_c=1.07$~eV. For an interaction strength of $U=1$~eV, the leading pairing has $s_{\pm}$-wave symmetry, and the associated gap function is shown in the lower-right panel of Fig.~\ref{fig:fermisurface}. Although the RPA is not quantitatively controlled in the strongly correlated regime, it provides a useful framework for identifying the leading pairing symmetry and the dominant momentum structure of the pairing interaction. A fully quantitative determination of $T_c$ and pairing strength is beyond the scope of this work.




{\it Discussion}. 
The recent discovery of nickelate superconductivity has provided a new platform for studying of high-temperature superconductivity. Their distinctive bilayer structure and pairing mechanism, which differ from those of cuprates and iron-based superconductors, further enrich the landscape of high-$T_c$ superconductivity. In contrast, high-temperature superconductivity has not yet been established in cobalt-based systems, leaving an important gap in the family of correlated 3$d$ superconductors. In this work, we predict a class of Co-based candidates that are structurally analogous to LNO and exhibit similarly strong electronic correlations. Our results suggest that pressure and electron doping can tune these cobaltates into a regime favorable for superconductivity, motivating future experimental synthesis and characterization.


We also construct bilayer two-orbital models for these Co-based materials and find, within the RPA, a leading $s$-wave pairing channel. It will be valuable to examine these models using complementary theoretical approaches beyond RPA, and to pursue a quantitative assessment of pairing strength and $T_c$.

Given that bulk bilayer nickelates often require pressure to become superconducting and tend to exhibit reduced crystal symmetry and competing density--wave orders, we expect that the cobalt-based materials may behave similarly at ambient pressure; this will be investigated in future work.

Building on our results, additional Co-based superconducting candidates are worth exploring. One direction is to replace La in LCO-Th with other trivalent rare-earth elements, for which many choices are available. Alternative routes to electron doping could also be pursued, such as hydrogen doping. For example, Ref.~\cite{yd8w-frs8} reported hydrogen doping in La$_2$NiO$_4$ with the aim of achieving electron-doped, single-orbital superconductivity in a nickel-based system. Like the cuprates and iron-based superconductors---which comprise large material families with diverse compositions and a wide range of $T_c$ values—--cobaltates may likewise form an extended class capable of hosting high-temperature superconductivity. Furthermore, the recent discovery of pressure-induced superconductivity in the trilayer Ruddlesden--Popper nickelate La$_4$Ni$_3$O$_{10}$ (with $T_c$ up to 30 K) opens another promising avenue for designing Co-based analogs. In La$_4$Ni$_3$O$_{10}$, the Ni ions have an average $d$-electron count corresponding to 3$d^{7.33}$, a filling that may be more naturally accessible in cobalt (which typically spans $3d^{6}$–$3d^{8}$) than in nickel, potentially facilitating analogous electronic correlations and superconducting instabilities in the corresponding cobaltates.

The research strategy employed here---leveraging structural analogy, electron doping, and theoretical modeling of multi-orbital correlations to predict superconductivity---does not straightforwardly extended to a cuprate analog of pressurized LNO. Previous theoretical studies~\cite{wang2025correlated,she2025absence} suggest that cuprates do not support two-orbital strong correlations of the type found in LNO, where both Ni $d_{x^2-y^2}$ and $d_{z^2}$ orbitals play active roles in driving the pairing instability. In cuprates, superconductivity is predominantly single-orbital in nature, centered on the Cu $d_{x^2-y^2}$ orbital within a Mott-insulating framework, which precludes genuine two-orbital high-$T_c$ superconductivity of the kind discussed for bilayer nickelates.

In contrast, iron-based superconductors are well-established to host multi-orbital (typically five-orbital) high-$T_c$ superconductivity, with Hund's coupling and orbital-selective correlations playing key roles. Given that cobalt is adjacent to iron in the periodic table and shares similar 3$d$ electronic configurations with tunable spin states, designing Co-based superconductors by analogy with iron-based systems (e.g., through appropriate structural motifs, doping, or pressure) offers a distinct and promising pathway toward cobalt-based high-temperature superconductivity.




\begin{acknowledgments}
This work was supported by the National Key R\&D Program of China (Grants No. 2024YFA1408601 and No. 2024YFA1408602) and the National Natural Science Foundation of China (Grant No. 12434009). J.X.W. was also supported by the Outstanding Innovative Talents Cultivation Funded Programs 2025 of Renmin University of China. Z.Y.L. was also supported by the Innovation Program for Quantum Science and Technology (Grant No. 2021ZD0302402). Computational resources were provided by the Physical Laboratory of High Performance Computing in Renmin University of China.
\end{acknowledgments}

\bibliography{lco}
%

\end{document}


\title{Supplementary Information for: \\Prediction of several Co-based La$_3$Ni$_2$O$_7$-like superconducting materials}
\author{Jing-Xuan Wang}
\thanks{These authors contributed equally to this work.}
\affiliation{School of Physics and Beijing Key Laboratory of Opto-electronic Functional Materials $\&$ Micro-nano Devices, Renmin University of China, Beijing 100872, China}\affiliation{Key Laboratory of Quantum State Construction and Manipulation (Ministry of Education), Renmin University of China, Beijing 100872, China}

\author{Yi-Heng Tian}
\thanks{These authors contributed equally to this work.}
\affiliation{School of Physics and Beijing Key Laboratory of Opto-electronic Functional Materials $\&$ Micro-nano Devices, Renmin University of China, Beijing 100872, China}\affiliation{Key Laboratory of Quantum State Construction and Manipulation (Ministry of Education), Renmin University of China, Beijing 100872, China}

\author{Jian-Hong She}
\thanks{These authors contributed equally to this work.}
\affiliation{School of Physics and Beijing Key Laboratory of Opto-electronic Functional Materials $\&$ Micro-nano Devices, Renmin University of China, Beijing 100872, China}\affiliation{Key Laboratory of Quantum State Construction and Manipulation (Ministry of Education), Renmin University of China, Beijing 100872, China}

\author{Rong-Qiang He}\email{rqhe@ruc.edu.cn}\affiliation{School of Physics and Beijing Key Laboratory of Opto-electronic Functional Materials $\&$ Micro-nano Devices, Renmin University of China, Beijing 100872, China}\affiliation{Key Laboratory of Quantum State Construction and Manipulation (Ministry of Education), Renmin University of China, Beijing 100872, China}

\author{Zhong-Yi Lu}\email{zlu@ruc.edu.cn}\affiliation{School of Physics and Beijing Key Laboratory of Opto-electronic Functional Materials $\&$ Micro-nano Devices, Renmin University of China, Beijing 100872, China}\affiliation{Key Laboratory of Quantum State Construction and Manipulation (Ministry of Education), Renmin University of China, Beijing 100872, China}\affiliation{Hefei National Laboratory, Hefei 230088, China}

\date{\today}

\maketitle
\renewcommand{\thepage}{S\arabic{page}}  
\renewcommand{\thesection}{S\arabic{section}}   
\renewcommand{\thetable}{S\arabic{table}}   
\renewcommand{\thefigure}{S\arabic{figure}}
\renewcommand{\theequation}{S\arabic{equation}}
\section{The DFT+DMFT method}


\begin{figure*}[htb]
  \centering
  \includegraphics[width=17.2cm]{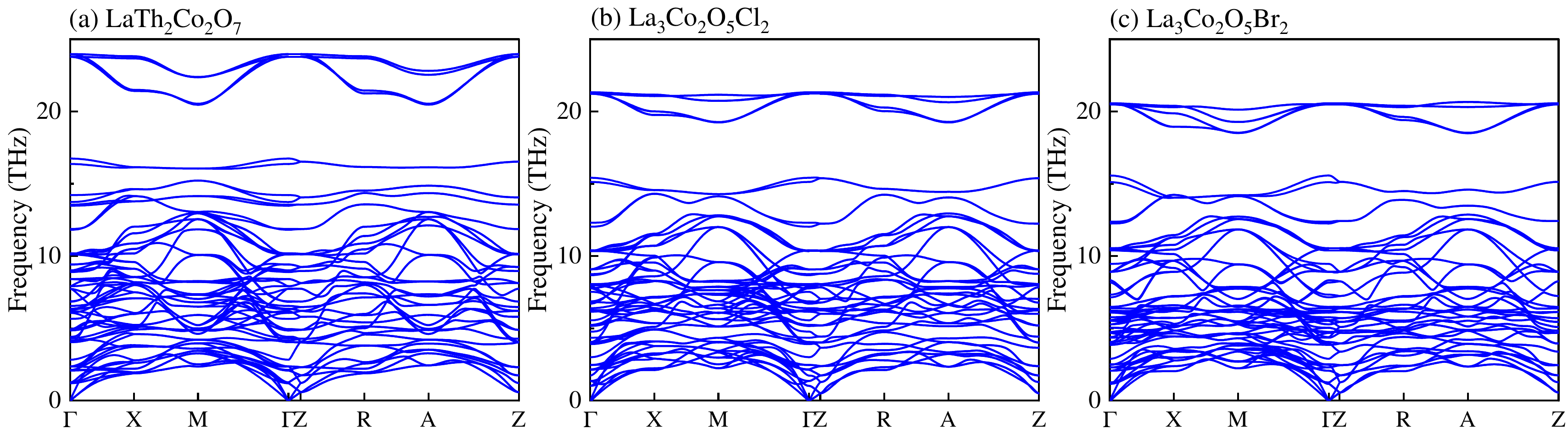}
  \caption{The calculated phonon spectra for (a) LCO-Th, (b) LCO-Cl, and (c) LCO-Br with 3 $\times$ 3 $\times$ 1 superstructure by DFT.}
  \label{fig:phonopy}
\end{figure*}

We preformed density functional theory (DFT) calculations using the VASP software and fully charge self-consistent DFT plus  dynamical mean-field theory (DFT+DMFT) calculations using the eDMFT package, based on WIEN2K \cite{Blaha2020jcp}. The generalized gradient approximation (GGA) of the Perdew-Burke-Ernzerhof (PBE) functional was used to describe the exchange-correlation energy \cite{Perdew1996prl}. We used the VASP software to structural optimization. The energy cutoff of the plane wave basis was set to 550 eV. The energy and force convergence accuracy are $10^{-8}$ and 1 meV/atom. The nonmagnetic state of the La$_3$Co$_2$O$_7$ (LCO), LaTh$_2$Co$_2$O$_7$ (LCO-Th), La$_3$Co$_2$O$_5$Cl$_2$ (LCO-Cl), and La$_3$Co$_2$O$_5$Br$_2$ (LCO-Br) were calculated in the DMFT calculations. The Co-3$d$ orbitals split to fully occupied ${t_{2g}}$ orbitals and partially occupied ${e_g}$ orbitals due to octahedral crystal field effect. Consequently, we restrict our treatment of correlation effects to the two ${e_g}$ orbitals of Co (${d_{z^2}}$ and ${d_{x^2-y^2}}$). The Hubbard $U$ and Hund’s coupling $J_H$ are chosen to be 5.0 eV and 1.0 eV, respectively. We consider the paramagnetic state for DFT+DMFT calculation. The calculations were converged with $R_{\mathrm{MT}}K_{\max}=7.0$; the muffin-tin radii $R_{\mathrm{MT}}$ were 2.11, 1.88, 1.62, 2.32, and 2.23 bohr for La, Co, O, Th, and Cl, respectively. A $25\times25\times25$ $k$-point mesh was used to sample the first Brillouin zone, and a $64\times64\times64$ mesh was employed to calculate the density of states. An energy window from $-10$ to $10$~eV was used to construct localized Co $3d$ orbitals. We employed the continuous-time quantum Monte Carlo (CTQMC) impurity solver together with the exact double-counting scheme~\cite{Gull2011rmp,Haule2015prl}. The number of Monte Carlo steps was set to $3\times10^{7}$. The real-frequency self-energy was then obtained by analytical continuation with the maximum entropy methods~\cite{Gull2011rmp}.


\section{Crystal structure details}

To facilitate comparison with the high-pressure phase of the La$_3$Ni$_2$O$_7$ (LNO) superconductor at 29.5 GPa, we applied a pressure of 25 GPa to the bulk phase of LCO, yielding similar lattice constants as listed in Table~\ref{tabS1}. In addition, the bond lengths and bond angles in LNO and doped LCO also are comparable. To confirm the stability of LCO with Th, Cl, or Br doped, we have calculated their phonon spectra with 3 $\times$ 3 $\times$ 1 superstructure using DFT. As shown in Fig.~\ref{fig:phonopy}, there is no obvious soft phonon mode, indicating that they are very stable.



\begin{table}[t]
  \centering
  \renewcommand\arraystretch{1.3}
  \caption{The lattice constants, bond lengths between transition metals M and O atoms, and the bond angles within the plane are shown for LNO$^\ast$ at 29.5 GPa, and LNO, LCO, LCO-Th, LCO-Cl, and LCO-Br at 25 GPa. The O1, O2, and O3 denote the in-plane O, the inner apical O, and the outer apical O, respectively.}
  \label{tabS1}
  \begin{tabular*}{8.6cm}{@{\extracolsep{\fill}} lcccccc}
  \hline\hline
    & \multicolumn{2}{c}{lattice constant} 
    & \multicolumn{3}{c}{bond length} 
    & \multicolumn{1}{c}{bond angle} \\
    \cmidrule(lr){2-3}\cmidrule(lr){4-6}\cmidrule(lr){7-7}
     & $a=b$ & $c$ & M-O1 & M-O2 & M-O3 & O1-M-O1\\
    \hline
    LNO$^\ast$& 3.680 & 19.362 & 1.841 & 1.901 & 2.048 &176.2\\
    LNO  & 3.699 & 19.482 & 1.850 & 1.908 & 2.066 &176.7\\
    LCO & 3.703 & 19.174 & 1.852 & 1.895 & 1.967& 176.5\\
    LCO-Th& 3.696 & 19.491 & 1.849 &1.963 &2.093 &175.9 \\
    LCO-Cl & 3.752 & 21.436 & 1.880 & 1.966 & 2.567 & 172.8 \\
    LCO-Br &3.770 & 22.197& 1.890 &1.971 & 2.718 & 171.9\\
    \hline\hline
  \end{tabular*}
\end{table}

\section{\label{appB} Density of states}

\begin{figure}[htb]
  \centering
  \includegraphics[width=8.6cm]{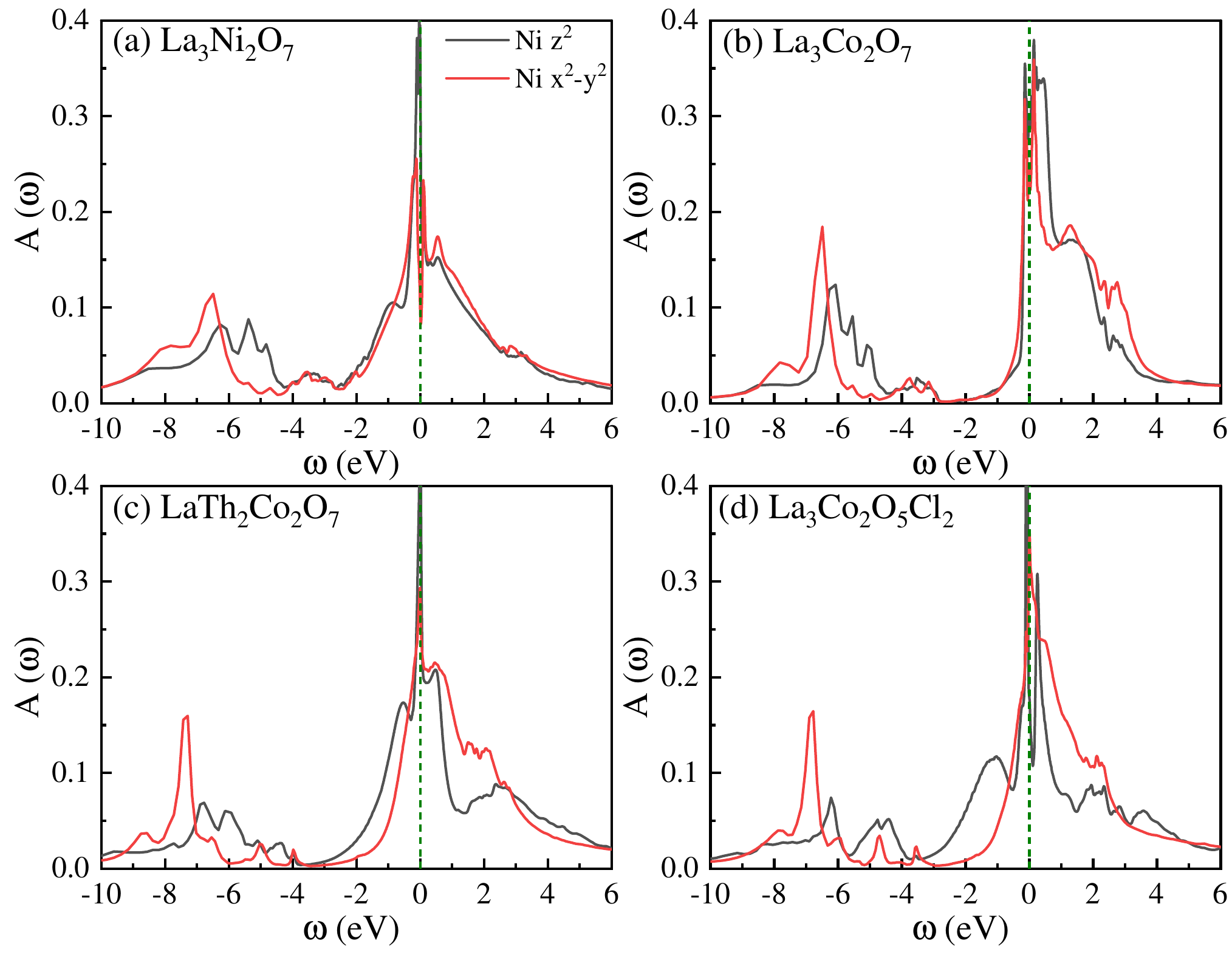}
  \caption{Orbital-resolved local spectral functions $A(\omega)$ for the ${d_{z^2}}$ (black) and $d_{x^2-y^2}$ (red) orbitals in (a) LNO, (b) LCO, (c) LCO-Th, and (d) LCO-Cl at high pressure.}
  \label{fig:dos}
\end{figure}

Fig.~\ref{fig:dos} displays the spectral function $A(\omega)$ of the $eg$ orbital for (a) LNO (b) LCO, and the electron-doped cobaltates (c) LCO-Th and (d)LCO-Cl. Compared to undoped LCO, the Fermi level of the spectral function for the doped compounds shifts to the right in Figs.~\ref{fig:dos} (c) and (d). The spectral function of the eg orbital in LCO-Th and LCO-Cl exhibit a narrow quasiparticle peak at the fermi level and two broad incoherent shoulders above and below the Fermi level. This reconstruction of the Co $eg$ manifold upon electron doping drives LCO from a pseudogap like state to a two orbital correlated metal whose density of states near $E_F$ is essentially the same as that of the nickelate LNO.


\section{Tight-binding models}

To study the electronic properties on LCO-Th and LCO-Cl, we construct tight-binding (TB) models that contains only Co-e$_g$ orbitals. The TB Hamiltonian is
\begin{equation}
\begin{aligned}
H_0&=\sum_{\rm{k} \sigma} \Psi_{\rm{k} \sigma}^{\dagger}  H(\rm{k}) \Psi_{\rm{k} \sigma},
\end{aligned}\label{eq:h}
\end{equation}
where $\Psi_{\sigma} = (d_{Ax\sigma}, d_{Az\sigma}, d_{Bx\sigma}, d_{Bz\sigma})^{T}$ and $d_{s\sigma}$ annihilates an $s = Ax, Az, Bx, Bz$ electron with spin $\sigma$. $A$ and $B$ label the top layer and the bottom layer, $x$ and $z$ label $d_{x^2-y^2}$ and $d_{3z^2-r^2}$ orbitals, respectively. The matrix $H(k)$ has the following form:
\begin{align}
H & ({\rm k})=\left(\begin{array}{cc}
H_{A}({\rm k}) & H_{AB}({\rm k})\\
H_{AB}({\rm k}) & H_{A}({\rm k})
\end{array}\right),\nonumber \\
H_{A}({\rm k})= & \left(\begin{array}{cc}
T_{{\rm k}}^{x} & V_{{\rm k}}\\
V_{{\rm k}} & T_{{\rm k}}^{z}
\end{array}\right),\qquad H_{AB}({\rm k})=\left(\begin{array}{cc}
T_{\mathbf{k}}^{'x} & V_{{\rm k}}^{\prime}\\
V_{{\rm k}}^{\prime} &T_{\mathbf{k}}^{'z}
\end{array}\right).\label{eq:tb}
\end{align}
with
\begin{equation}
\begin{aligned}
T_{\mathbf{k}}^{x/z}
&= 2t_{1}^{x/z}(\cos k_x + \cos k_y) 
  + 4t_{2}^{x/z}\cos k_x\cos k_y \\
 &+ 2t_{5}^{x/z}(\cos 2k_x + \cos 2k_y)+ \epsilon_{x/z}, \\
V_{\mathbf{k}}
&= 2t_{3}^{xz}(\cos k_x - \cos k_y)
  + 2t_{5}^{xz}(\cos 2k_x - \cos 2k_y), \\
T_{\mathbf{k}}^{'x/z}
&= t_{\perp}^{x/z} + 2t_{3}^{x/z}(\cos k_x + \cos k_y), \\
&+ 4t_{4}^{x/z}\cos k_x\cos k_y+ 2t_{6}^{x/z}(\cos 2k_x + \cos 2k_y), \\
V'_{\mathbf{k}} 
&= 2t_{4}^{xz}(\cos k_x - \cos k_y).
\end{aligned}
\end{equation}


Due to the very close energetic proximity between the La/Th$f$ and Co-3$d$ levels, the disentanglement of Wannier downfolding are unreliable. So, the parameters for the tight-binding model presented above were determined via a least-squares fit to the DFT bands. To perform this fit, we first selected a subset of k-points from the $\Gamma$-$X$-$M$-$\Gamma$ path where the bands possess strong $e_g$ orbital character, and then fit our model to the DFT energies at only these points.

\section{RPA Approach}

To investigate the magnetic and superconducting properties, we employ the multi-orbital Random Phase Approximation (RPA) to account for electron-electron interactions. The interaction Hamiltonian is given by:
\begin{equation}
\begin{aligned}
H_I&=U\sum_{il\alpha} n_{il\alpha\uparrow}n_{il\alpha\downarrow} \\
&+ \sum_{il\sigma \sigma'}(U^{\prime}-\delta_{\sigma \sigma'}J) n_{ilx\sigma} n_{ilz\sigma'} \\
&-J\sum_{il} (d_{ilx\uparrow}^\dagger d_{ilx\downarrow} d_{ilz\downarrow}^\dagger d_{ilz\uparrow} +\mathrm{h.c.})\\
&+J\sum_{il}(d_{ilx\uparrow}^\dagger d_{ilx\downarrow}^\dagger d_{ilz\downarrow} d_{ilz\uparrow} +\mathrm{h.c.})
\end{aligned}\label{eq:h}
\end{equation}
where $U$, $U^{\prime}$, and $J$ represent the intra-orbital interaction, inter-orbital interaction, and Hund's coupling, respectively. Following standard convention, we enforce rotational invariance by setting $U^{\prime} = U - 2J$ and fix the ratio $J = U/6$.

Within the RPA formalism, the renormalized spin ($\chi^{(s)}$) and charge ($\chi^{(c)}$) susceptibilities are calculated from the bare susceptibility tensor $\chi^{(0)}$ as:
\begin{equation}
\begin{aligned}
\chi^{(s)}(\mathbf{k}, i\omega)
&= \bigl[I - \chi^{(0)}(\mathbf{k}, i\omega)\,U^{(s)}\bigr]^{-1}
  \chi^{(0)}(\mathbf{k}, i\omega),\\[4pt]
\chi^{(c)}(\mathbf{k}, i\omega)
&= \bigl[I + \chi^{(0)}(\mathbf{k}, i\omega)\,U^{(c)}\bigr]^{-1}
  \chi^{(0)}(\mathbf{k}, i\omega).
\end{aligned}
\label{eq:chi_sc}
\end{equation}
Here, $U^{(s)}$ and $U^{(c)}$ denote the interaction tensors in the spin and charge channels, respectively.

We utilize these renormalized susceptibilities to construct the effective pairing interaction mediated by spin fluctuations. By treating the interaction at the mean-field level, we derive a linearized gap equation. Solving the eigenvalue problem of this equation yields the pairing strength, represented by the eigenvalue $\lambda$, and the corresponding gap symmetry function $g(k)$. For further details on the multi-orbital RPA formalism, we refer to Refs.~\cite{graserNJP2009,kemperNJP2010,zhangPRB2024}.

In our calculations, we considered both singlet and triplet pairing channels. However, we found that the leading eigenvalues for singlet pairing are significantly larger than those for triplet pairing in both materials, indicating a dominant singlet pairing state.



\section{Details of the RPA Calculation}
\begin{figure}[htbp!]
  \includegraphics[width=8.6cm]{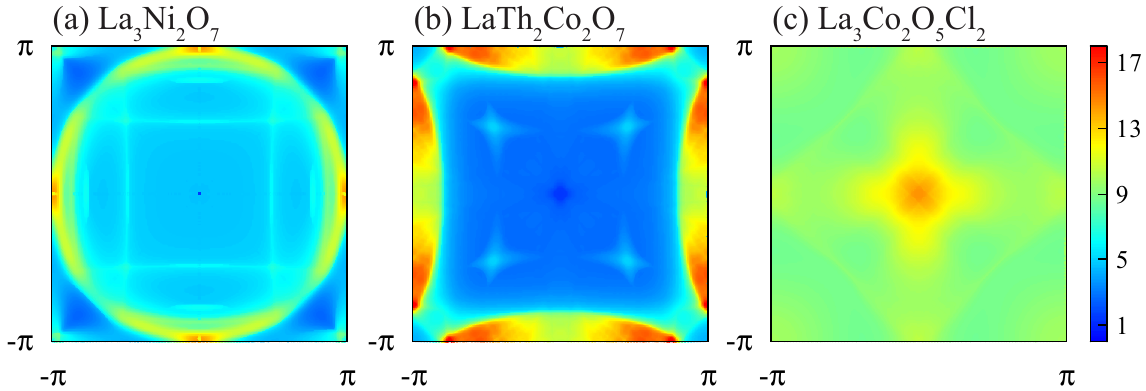}
  \caption{Distribution of the leading eigenvalue of the static spin susceptibility matrix $\chi^{(s)}$, in the Brillouin zone.The calculations are shown for our TB model with specific interaction strengths for each material (a) LNO at $U=1.4$ eV, (b) LCO-Th at $U=0.9$ eV, and (c) LCO-Cl at $U=1$ eV.}\label{fig:chis}
\end{figure}

\begin{figure}[htbp!]
  \includegraphics[width=8.6cm]{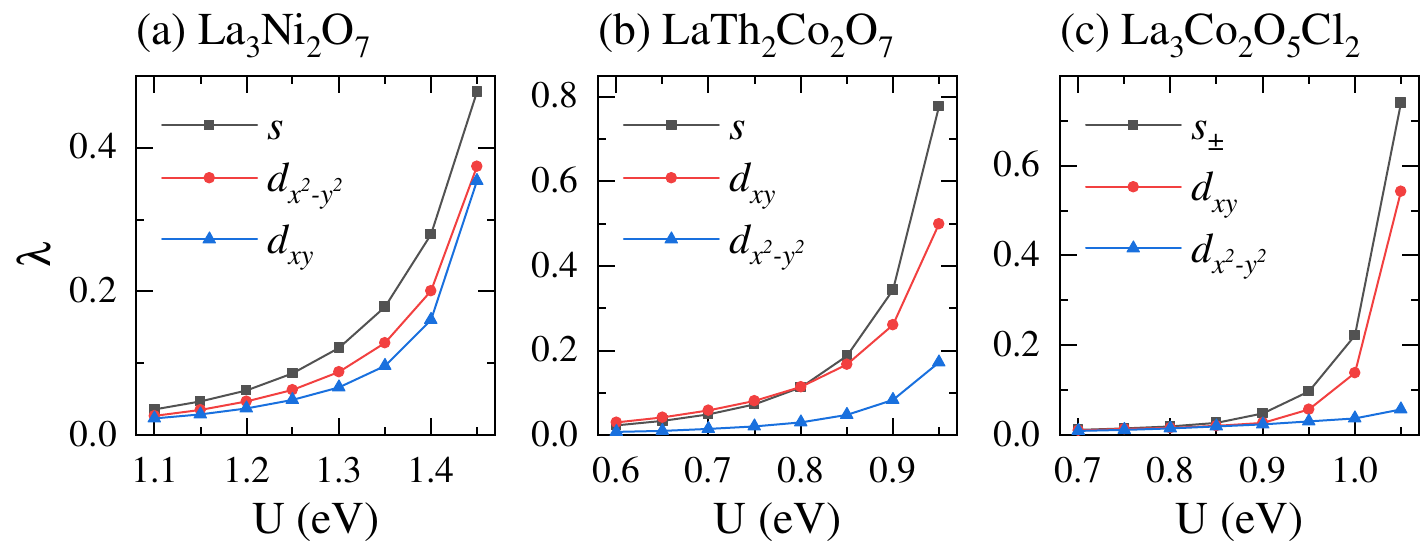}
  \caption{The leading pairing eigenvalue $\lambda$, obtained from RPA calculations for our TB model. The figure displays $\lambda$ as a function of U for various pairing symmetries. A comparison is made between (a) LNO, (b) LCO-Th, and (c) LCO-Cl.}
  \label{fig:eigen}
\end{figure}

Fig.~\ref{fig:chis} shows the largest eigenvalue of the RPA static spin susceptibility matrix $\chi^{(s)}(q)$ for each material. Fig.~\ref{fig:eigen} illustrates how the pairing eigenvalues $\lambda$ for three dominant pairing symmetries grow as the interaction $U$ increases for each material. The leading pairing symmetry is found to be the $s$-wave for all three materials as $U$ approaches the critical value. 

\bibliography{lco}